\documentclass[12pt,fleqn]{article}
\usepackage[american]{babel}

\def\beq{\begin{equation}}
\def\eeq{\end{equation}}
\def\6{\langle}
\def\9{\rangle}
\def\tr{\mbox{tr}\,}
\def\half{\mbox{$1\over2$}\,}

\begin{document}

\vspace*{\fill}
\begin{center}
{\Large {\bf Hybrid classical-quantum dynamics}}\\[15mm]

Asher Peres and Daniel R. Terno $^*$\\[8mm]
{\sl Department of Physics, Technion---Israel Institute of Technology,
32\,000 Haifa, Israel}\\[15mm]

{\bf Abstract}

\end{center}

A hybrid formalism is proposed for interacting classical and quantum
sytems. This formalism is mathematically consistent and reduces to
standard classical and quantum mechanics in the case of no interaction.
However, in the presence of interaction, the correspondence principle is
violated.\vfill

 \vfill

\noindent $^*$ Electronic mail: \
{\tt peres@photon.technion.ac.il, terno@physics.technion.ac.il}
\newpage

Quantum mechanics gives exceedingly accurate predictions for atomic and
nuclear systems. Classical mechanics is just as successful for planetary
motion. Can there be a theory encompassing both classical and quantum
mechanics, so that atoms and planets can be treated in a unified way?
Indeed, it was speculated long ago that the gravitational field is
classical, even though its material sources are quantized~\cite{moller}.
Niels Bohr always insisted that measuring instruments must be described
in classical terms~\cite{bohr}. However, Bohr did not provide a dynamical
description of their interaction with quantum systems.  A {\it dual\/}
(classical and quantum) description of measuring instruments is
possible~\cite{hay}, but this is not a truly {\it hybrid\/} formalism
as proposed below, because in that dual formalism the interaction is
always between subsystems of the same kind.

If we try to write equations of motion that combine classical canonical
variables and quantum operators, we find that the classical variables,
whose values ought to be ordinary numbers, pick up operator components,
which is manifestly unacceptable (we shall later show how to overcome
this difficulty). There have been numerous claims and counter-claims
as to the possibility of existence a consistent classical-quantum
formalism.  Some recent articles are listed in the
references~\cite{claims}. There is no real contradiction between them,
because their conflicting conclusions result from different demands
for consistency.

In this article, we propose a hybrid formalism for interacting classical
and quantum sytems. This formalism is mathematically consistent and
reduces to standard classical and quantum mechanics if there is no
interaction between the systems. However, in the presence of interaction,
the correspondence principle is violated, so that this hybrid formalism
appears physically abnormal.

Our work follows a remark by Jauch~\cite{jauch} that if we assume that
not all quantum dynamical variables are actually observable, and if we
set rules for distinguishing measurable from nonmeasurable operators,
it is then possible to define a classical system as a special type
of quantum system for which all measurable operators commute. All
the observable dynamical variables of such a classical system can be
simultaneously diagonalized and therefore can, in principle, have
sharp values. If we are confined to this set of observables and we
use a basis where they all are diagonal, then only diagonal elements of
density matrices can be reconstructed from measurement statistics and
have an operational meaning. Quantum states (i.e., density matrices)
with the same diagonal elements are indistinguishable. We may therefore
consider all states as pure by {\it defining\/} their off-diagonal
elements as $\rho_{mn}=\sqrt{\rho_{mm}\rho_{nn}}$, so that $\rho$ always
has rank~1. It is then possible to define a classical state vector $\psi$
by $\rho=\psi\psi^\dagger$.

Pure states defined in this way do not imply that dynamical variables
have sharp values. The situation is just as in classical statistical
mechanics: the expectation value of an observable $A$ is given by

\beq \6A\9=\tr(\rho A), \eeq
and we may then have $\6A^2\9>\6A\9^2$, as usual. Note that such a
``classical'' system also has noncommuting operators, but the latter
should be considered as abstract mathematical expressions which are
not experimentally observable. With this formal definition, the notion
``classical'' acquires a meaning with respect to a specified set of
dynamical variables. This does not preclude that, by using more elaborate
observation means, every physical system might display quantum features.

To completely mimic classical mechanics, we need a quantum algorithm
that reproduces exactly the equations of motion of a
classical canonical system. For this, we shall follow Koopman's
formalism~\cite{koopman,qt}. We consider for simplicity a single degree
of freedom and denote the canonical variables as $x$ and $y$ (rather
than $x$ and $p$, as usual, because we wish to reserve the symbol $p$
for the momentum of a quantum system, to be introduced later). Let us
write the Liouville equation as

\beq i\,\partial f/\partial t=Lf, \eeq
where $L$ is the Liouville operator, or Liouvillian,

\beq L=\left({\partial H\over\partial y}\right)
 \left(-i{\partial\over\partial x}\right)
 -\left({\partial H\over\partial x}\right)
 \left(-i{\partial\over\partial y}\right). \label{L} \eeq
The Liouville density $f$ is never negative. Since its quantum analog,
namely Wigner's distribution~\cite{qt,wigner} (which may be negative),
is quadratic in the quantum wave function, it is convenient to introduce
likewise a ``classical wave function'' $\psi=\sqrt{f}$, which in this
case satisfies the same equation of motion as $f$,

\beq i\,\partial\psi/\partial t=L\psi. \eeq

We shall now consider $\psi$ as the fundamental entity (but only
$f=|\psi|^2$ has a direct physical meaning). It can be proved that,
under reasonable assumptions about the Hamiltonian, the Liouvillian
is an essentially self-adjoint operator and generates a unitary
evolution~\cite{reed}:

\beq \6\psi,\phi\9:=\int\psi(x,y,t)^*\,\phi(x,y,t)\,dxdy={\rm const.}
\eeq
The introduction of a Hilbert space with unitary dynamics enables one
to use familiar methods of quantum mechanics for the analysis of ergodic
problems~\cite{reed} and of classical chaos~\cite{chaos}. It is possible
to further mimic quantum theory by introducing {\it commuting\/} operators
$\hat{x}$ and $\hat{y}$, defined by

\beq \hat{x}\,\psi=x\,\psi(x,y,t)\qquad{\rm and}
  \qquad\hat{y}\,\psi=y\,\psi(x,y,t). \eeq
Note that the momentum $\hat{y}$ is not the shift operator (the latter
is $\hat{p}_x=-id/dx$).  Likewise the boost operator is
$\hat{p}_y=-id/dy$. These two operators are not observable. We shall
henceforth omit the hats over the classical operators, as there is
no danger of confusion.

The analogy with quantum mechanics can be pushed further. What we have
above is a ``Schr\"odinger picture'' (operators are constant, wave
functions evolve in time as $\psi(t)=U(t)\psi(0)$, where $U(t)=e^{-iLt}$
if the Hamiltonian is time-independent). We can also define a ``Heisenberg
picture'' where wave functions are fixed and operators evolve:

\beq X_H(t)=U^\dagger XU.\eeq
The Heisenberg equation of motion,

\beq i\,dX_H/dt=[X_H,L_H]=U^\dagger[X,L]\,U, \eeq
together with the Liouvillian (\ref{L}), readily give Hamilton's
equations 

\beq {dx\over dt}={\partial H\over\partial y},\qquad\qquad
     {dy\over dt}=-{\partial H\over\partial x}.\eeq
There is however an important difference: the time translation operator
$L$ is not the energy, and its spectrum may extend to $-\infty$. For
example, if we have a harmonic oscillator with $H=(x^2+y^2)/2$, the
Liouvillian is

\beq L=y\,p_x-x\,p_y, \label{Lz}\eeq
whose eigenvalues are all the integers, negative as well as positive.
There is nothing wrong in that, since $L$ involves the unobservable
shift operators $p_x=-id/dx$ and $p_y=-id/dy$, and therefore $L$ itself
is not observable. Note that the solution of Hamilton's equations does
not introduce non-observable components into the observable variables,
because the Hamiltonian (contrary to the Liouvillian) involves only the
observables $x$ and $y$.

It is easy to introduce into the above dynamical formalism a quantum
system, with conjugate dynamical variables $q$ and $p$, as long as
the two systems do not interact. In the Schr\"odinger picture, we have
a wave function $\psi(x,y,q)$ whose evolution is given by

\beq \partial\psi/\partial t=K\psi, \eeq
where $K$ is ``Koopmanian'' 

\beq K=L(x,y,p_x,p_y)+H(q,p). \eeq
In that wave function, the coordinates $x$, $y$, and $q$ may be
entangled. In any case, we can obtain the reduced density matrices of
the ``classical'' and quantum systems by means of partial traces on the
other system, as usual.

The non-trivial problem is to introduce an interaction between the two
systems. An apparent difficulty occurs in the Heisenberg picture because
the equations of motion will mix variables of all kinds: classical
observables, non-observables, and quantum operators. This still is
acceptable, because we do not predict actual values for these variables,
but only expectation values, $\6A\9=\tr(\rho A)$, and no contradiction
may occur. Anyway, we can always use the Schr\"odinger picture, where
we know how to handle entangled wave functions.

The true difficulty, as we shall now see in a simple example, is that
the correspondence principle fails. Consider two harmonic oscillators,
with a bilinear coupling $kqx$, where $k$ is a constant. If we treat
both of them classically, with a Hamiltonian

\beq H=\half(q^2+p^2+x^2+y^2)+kqx, \label{H} \eeq
we obtain equations of motion

\beq \dot{q}=p,\qquad\qquad \dot{p}=-q-kx, \eeq
\beq \dot{x}=y,\qquad\qquad \dot{y}=-x-kq, \label{xy} \eeq
and there are two characteristic frequencies, $\omega=\sqrt{1\pm k}$,
corresponding to the normal modes $(q\pm x)$.  Exactly the same
equations of motion appear in the Heisenberg picture for quantum
mechanics. Likewise, in the Schr\"odinger picture, whenever there
is a quadratic potential, the differential equation for the Wigner
distribution is identical to the Liouville equation in classical
mechanics~\cite{qt}.  In view of this formal agreement of classical
and quantum mechanics, it is natural to demand that any hybridization
of the system in Eq.~(\ref{H}) shall also give the same equations
of motion. {\it This is the definite benchmark we propose for an
acceptable classical-quantum hybrid formalism.\/} It is not obvious
that this criterion can be achieved, because the hybrid system has no
normal modes $(q\pm x)$.

Let us try to obtain the above equations of motion from a Koopmanian

\beq K=\half(q^2+p^2)+(y\,p_x-x\,p_y)+K_i, \eeq
with a suitably chosen interaction term $K_i$. We cannot
have both $[p,K_i]=-kx$ and $[y,K_i]=-kq$ together with $[y,p]=0$,
because these equations are incompatible with Jacobi's identity

\beq [y,[p,K_i]]+[p,[K_i,y]]+[K_i,[y,p]]\equiv0. \eeq
The best result we were able to obtain was by ``koopmanizing'' the
interaction term $kqx$ in the Hamiltonian (\ref{H}). We have, from
Eq.~(\ref{L}),

\beq K_i=ikq\partial/\partial y\equiv -kqp_y. \eeq
The resulting equations of motion are Eq.~(\ref{xy}), unchanged, and

\beq \dot{q}=p,\qquad\qquad \dot{p}=-q-kp_y, \label{qp} \eeq
\beq \dot{p}_x=p_y,\qquad\qquad \dot{p}_y=-p_x. \eeq
The last equation is necessary, because an unobservable variable $p_y$
appears in Eq.~(\ref{qp}). The solution of (\theequation) for $p_y$ is
a superposition of $\sin t$ and $\cos t$ terms. When substituted into
Eq.~(\ref{qp}), these terms behave as a driving force with resonant
frequency, so that $q$ and $p$ include terms behaving as $t\sin t$
and $t\cos t$. The amplitude of the quantum oscillator increases
linearly with time, and energy is not conserved. It need not be: what
is conserved is the Koopmanian, which includes a term $(yp_x-xp_y)$
whose spectrum extends to $-\infty$.

In conclusion, there is no mathematical inconsistency in the hybrid
formalism that we proposed. However, it violates the correspondence
principle which we would expect to hold exactly for a pair of oscillators
with bilinear coupling. Therefore such a theory appears quite abnormal
from the point of view of physics.

\bigskip We thank Lajos Di\'osi for stimulating discussions.
DRT was supported by a grant from the Technion Graduate School. Work by
AP was supported by the Gerard Swope Fund and the Fund for Encouragement
of Research.

\end{document}